# Air-Scan-Based Point-Spread-Function Estimation for Megavoltage CT with Flat-Panel Detectors

**Seemeen Karimi, Joseph Bendahan, and K. Aditya Mohan**

Lawrence Livermore National Laboratory
7000 East Ave, Livermore, CA 94550
(617) 953-1662; karimi1@llnl.gov

## ABSTRACT

Accurate spatial-resolution characterization of flat-panel detectors at megavoltage energies is difficult due to long-tailed cross-talk. Conventional measurements at high energy using line-pair gauges or knife-edge phantoms require expensive, carefully fabricated objects and precise alignment, and despite these, scatter from the phantom, detector, and imaging room can further confound the measurement. We present a method for estimating the detector point spread function (PSF) from the mean noise power spectrum (NPS) calculated from repeated radiographs in air. The approach is mathematically equivalent to an autocovariance-based method but is more practical for long-tailed PSFs. We validate the method using simulated air data and Monte Carlo N-particle (MCNP6) calculations of energy deposition in a flat-panel detector. We show that deconvolving the experimental PSF increases the accuracy of transmission radiographs of step wedges. The method requires no phantom or alignments and is straightforward to reproduce.

**Keywords:** megavolt imaging, flat-panel detector, point spread function, noise power spectrum, cross-talk, deconvolution, MCNP

## INTRODUCTION

Megavoltage x-ray imaging systems are used in a wide range of applications, including radiotherapy portal imaging, nondestructive testing, and security inspection [1][2][3]. Flat-panel detectors are widely used for megavoltage radiography and computed tomography (CT). In these systems, spatial resolution is determined not only by the detector pixel size but also by cross-talk, i.e., secondary x-rays and optical photons depositing signal in pixels neighboring the primary pixel. At MeV energies, the cross-talk produces a PSF with low-amplitude tails with large support that integrates to a non-negligible contribution. Consequently, cross-talk not only blurs edges but also causes errors in measured transmission. The measured transmission errors are largest for small dense radiographed objects because long PSF tails transfer signal from the surrounding high-count air region into the low-count region of the object.

Measuring the PSF accurately at MV energies is more challenging than at energies below a few hundred kV. Line-pair gauges become ineffective at high spatial frequencies because thin bars have insufficient attenuation [4]. Thicker bars are challenging and expensive to fabricate, and extremely sensitivity to alignment. Knife-edge methods [5], including tungsten roll-bar measurements, are less sensitive to alignment than line-pair methods, but remain susceptible to scatter from the edge object and the surrounding room. They also require a large, precisely fabricated absorber. The roll-bar has previously been used to characterize source blur using film [6].

This work shows an alternative method to estimate PSF based on repeated air scans. The method relies on noise to estimate the detector transfer function. We had previously published an air scan approach that computes the PSF through the computation of autocorrelation [7] but the computations involving large covariance matrices needed for long-tailed functions become impractical. The method here involves calculation of DFTs, and is theoretically a mathematical equivalent of our earlier method, but practically involves different design choices. Our experimental system was a 9-MeV x-ray imaging system with a 3-mm tantalum filter, an ~1-mm focal spot and a magnification of ~5/6. Our detector was a Varex 1621AN flat-panel detector with a DRZ+ gadolinium oxysulfide scintillator and nominal 0.2-mm pixels.

## THEORY

Consider an ideal x-ray source with expected value of fluence $q$, which is constant everywhere. Let the detector be modeled as a linear, shift-invariant system with a PSF $h(\boldsymbol{r})$ where $\boldsymbol{r} = (r_x, r_y)$ represents spatial coordinates. The measured image is:

$$g(\boldsymbol{r}) = h(\boldsymbol{r}) * [q + n(\boldsymbol{r})], \qquad (1)$$

where * denotes convolution, and $n$ represents noise. Since the mean of the input field is constant, the mean value $h(\boldsymbol{r}) * q \to q$ can be extracted from measured data and subtracted. Then the detector response can be represented in the Fourier domain as

$$g'(\boldsymbol{r}) = h(\boldsymbol{r}) * n(\boldsymbol{r}), \qquad (2)$$

so that

$$G'(\mathbf{k}) = H(\mathbf{k})N(\mathbf{k}), \tag{3}$$

where $H(\mathbf{k})$ is the modulation transfer function (MTF) and $\mathbf{k} = (k_x, k_y)$ is spatial frequency. We assume that x-rays incident on the detector are independent and identically distributed (IID) according to a Poisson distribution function. Therefore they have a white noise power spectrum (NPS). If the detector is linear and stationary, the output NPS is therefore

$$S_g(\mathbf{k}) = S_n|H(\mathbf{k})|^2, \tag{4}$$

where $S_n$ is the input quantum NPS. This relation is the basis of this method. Any frequency dependence in the normalized NPS is attributed to detector blur, subject to the assumptions of independent incident quanta, adequate gain correction, and spatial stationarity. The equations above give us a digital filter. For a rectangular pixel with dimensions $\Delta_x$ and $\Delta_y$, the pixel-aperture transfer function is

$$P(k_x, k_y) = \operatorname{sinc}\left(\frac{k_x\Delta_x}{2}\right)\operatorname{sinc}\left(\frac{k_y\Delta_y}{2}\right), \tag{5}$$

The measured transfer function includes both detector blur and pixel integration:

$$H_{\text{det}}(\mathbf{k}) = H_{\text{meas}}(\mathbf{k})P(\mathbf{k}). \tag{6}$$

# MATERIALS AND METHODS

## Implementation Details

A set of 300 air scans was acquired and corrected for the dark measurement. Let $A_m(\mathbf{r})$ denote the dark-corrected image from scan $m$. Each image $A_m(\mathbf{r})$ comprises noise and the spatially varying gain of the detector. Each image was normalized by its mean $\bar{A}(\mathbf{r})$, where the mean is computed per pixel over time ($m$), and one is subtracted to yield a zero mean noise realization as follows:

$$x_m(\mathbf{r}) = \frac{A_m(\mathbf{r})}{\bar{A}(\mathbf{r})} - 1. \tag{7}$$

Gain normalization is essential. Without it, fixed pixel-to-pixel gain differences appear as spatial structure in the noise data and contaminate the estimated PSF. $x_m(\mathbf{r})$ in Equation (7) is similar to $g'(\boldsymbol{r})$ in Equation (2). Now, its DFT was calculated:

$$X_m(\mathbf{k}) = \mathcal{F}\{x_m(\mathbf{r})\}. \tag{8}$$

The NPS estimate was formed by averaging the squared magnitude:

$$H_{meas}(\mathbf{k}) = \frac{1}{M}\sum_{m=1}^{M}|X_m(\mathbf{k})|^2 \tag{9}$$

The DC value of this NPS was forced to be zero by the subtraction of one in Equation (7). Therefore, the DC value for the PSF was estimated by fitting a 3$^{rd}$-order polynomial to the first few low-frequency MTF components and extrapolating to zero frequency. The pixel-aperture response was included using the two-dimensional sinc factor. The PSF was obtained through an inverse Fourier transform. The resulting PSF was fitted with a Voight curve for use in image deconvolution.

## Verification and validation with simulations

The method was first tested using simulated air data. Poisson-distributed noise realizations were generated with spatially varying means to represent detector gain variation. The data were convolved with a Voight-shaped blurring kernel. Using our method, we extracted the PSF and found it to be a match to the blurring kernel. A second validation used MCNP6 simulations of detector energy deposition. Detailed detector information, including scintillator and electronic component geometry, was available from the detector manufacturer under a non-disclosure agreement. A pencil beam was incident on the detector, and deposited energy was calculated in annuli for 55 discrete photon energies and 54 radial locations. The simulation represented detector cross-talk but excluded room scatter, source variation, spectrum fluctuations, variable gain, and other experimental imperfections. Energies and annuli were unevenly sampled, to improve estimates from photon statistics as shown in Figure 1 (a).

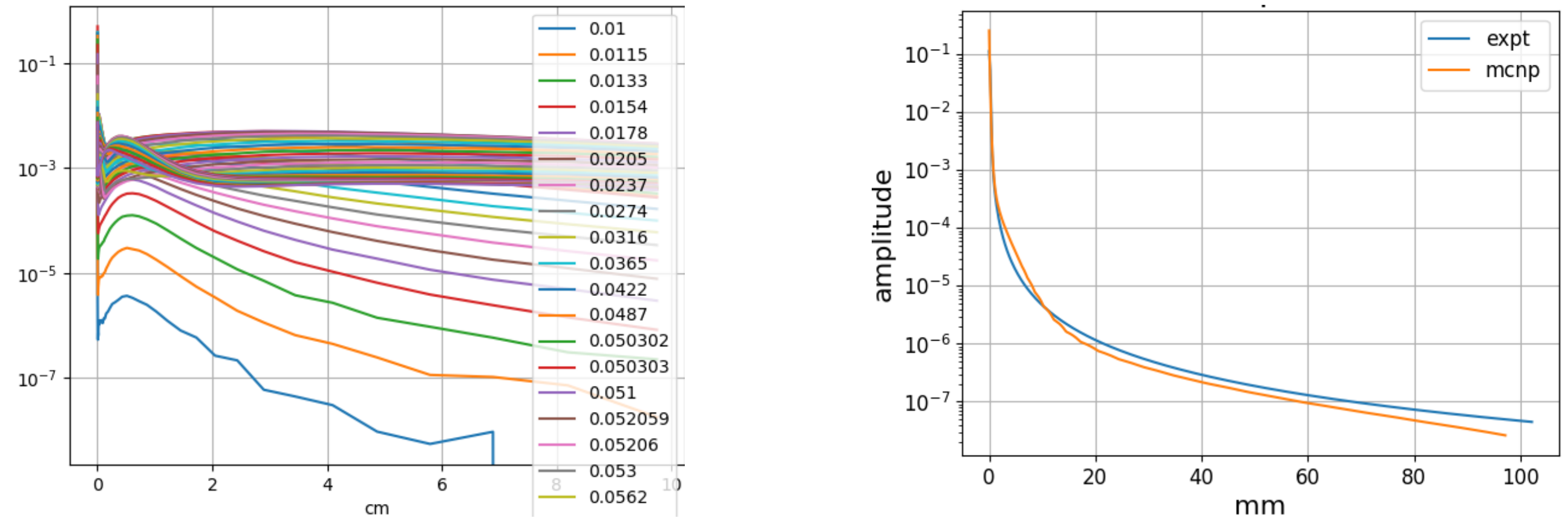

(a) (b)

**Figure 1. MCNP6 simulation of energy deposition in annular rings (a). Experimental and MCNP PSFs (b).**

The MCNP energy-deposition data were combined with the estimated scanner spectrum. If $D_i(r)$ is the radial deposited-energy distribution for energy $E_i$ and $w_i$ is the corresponding spectral weight, the spectrum-weighted cumulative deposition can be written as

$$C(r) = \sum_i w_i \int_0^r D_i\,(r')\,dr'.$$

To obtain a finely sampled radial distribution, the cumulative function was resampled at approximately 0.01-mm spacing. Finite differences were taken to give energy deposited in resampled annuli and then radial deposition density was calculated:

$$\rho_j = \frac{C(r_{j+1}) - C(r_j)}{\pi\left(r_{j+1}^2 - r_j^2\right)}.$$

The radial density $\rho_j$ was mapped to a two-dimensional, radially symmetric PSF on a grid, with downsampling by area averaging to the detector pixel pitch, yielding a 2-D PSF on a rectangular grid.

## RESULTS

## Experimental and simulated PSFs

The MCNP-derived PSF exhibited a narrow central response accompanied by long, low-amplitude tails. This behavior is expected for a high-energy scintillator detector, where secondary interactions transport energy over distances substantially larger than the nominal pixel pitch. The MCNP calculation was useful for interpreting the experimental result. The experimental and MCNP PSFs are shown in Figure 1 (b). The differences may include imperfect knowledge of detector construction, the assumption of perfect stationarity, fitting the DC term, and insufficient accuracy in the experiment, given the large dynamic range of the problem.

## Step-wedge measurements

The effect of the PSF tails was evident in radiographs of step wedges as shown by profiles in Figure 2. For a small tungsten wedge, most of the detector area was exposed to x-rays through air. The high-count air region therefore contributed signal through the long PSF tails into the wedge-shadow region. The uncompensated radiograph (labeled “Original”) consequently showed a transmission bias compared to the ground truth (“GT”) derived with the help of XCOM tables. Richardson–Lucy deconvolution [9] using the experimentally derived PSF (“Deconv:expt”) increased quantitative agreement. Deconvolution using the MCNP-derived PSF produced a similar result (“Deconv: mcnp”). These results indicate that correcting the detector PSF can improve not only visual sharpness but also quantitative transmission accuracy. The remaining discrepancy with ground truth may be due to scatter, source blur, and errors in the experimental PSF. The stainless-steel wedge showed improved flattening after deconvolution and closer agreement with ground truth, particularly in the low-attenuation region. This wedge was a larger one than the tungsten wedge, and radiographed with collimation. Therefore, the discrepancy with GT is smaller to begin with.

## 5. DISCUSSION

The air-scan method eliminates the need for a large, expensive phantom and avoids the alignment sensitivity associated with line-pair and knife-edge measurements. Air scans are easy to repeat, allowing the measurement to be incorporated into detector acceptance testing, periodic quality assurance, or system characterization after detector replacement. However, there are limitations. First, the method assumes wide-sense stationarity. Second, the method can only estimate detector blur, not source or system blur. Third, room scatter is not necessarily distinguishable from detector cross-talk in an air-scan measurement. Scatter generated in the room or by nearby structures can create broad spatial correlations. Collimation, shielding, and measurements at multiple source-to-detector configurations can help separate these contributions. Fourth, aliasing is inherent in a sampled detector,which biases the PSF. The method also remains less accurate over some intermediate spatial range, as evidenced by the residual slopes of the steps. Comparing the present estimator with the covariance method is an important next step for quantifying these errors.

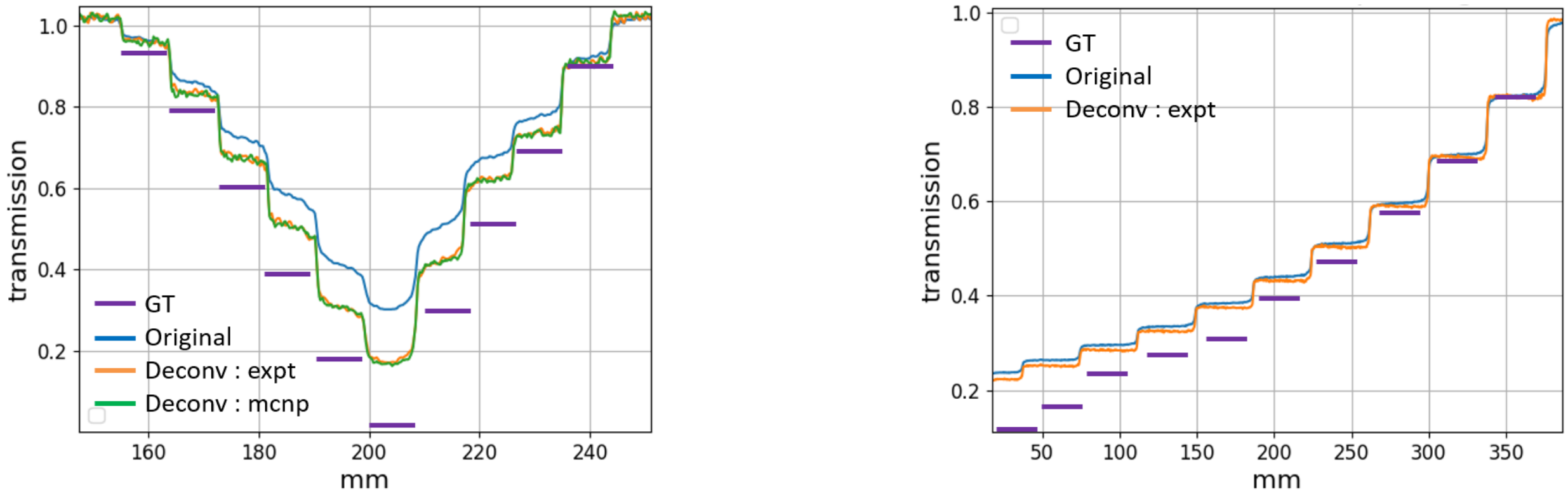

**Figure 2. Transmission radiographs of a tungsten step wedge (left), and a stainless steel step wedge (right).**

# 6. CONCLUSION

A practical phantom-free method was demonstrated for estimating the PSF of a flat-panel detector from repeated air scans. The method uses gain-normalized noise images, direct DFT-based NPS estimation, pixel-aperture correction, and inverse Fourier transformation. It is theoretically equivalent to an autocovariance-based estimator but is more convenient for detectors with broad PSF support. Deconvolving the measured PSFs improved transmission accuracy of transmission images relative to uncompensated measurements. Its principal limitations are the assumption of stationarity, incomplete treatment of source blur and room scatter, aliasing, and reduced accuracy over portions of the intermediate-frequency range. Future work will compare the DFT estimator quantitatively with the covariance method and develop spatially resolved and source-blur-aware extensions.

## Acknowledgments

The authors acknowledge prior MCNP detector modeling that provided the detailed detector description and energy-deposition data used in this study. This work was performed under the auspices of the U.S. Department of Energy by Lawrence Livermore National Laboratory under contract DE-AC52-07NA27344 and LDRD 25-ERD-017. This document is released as LLNL-PROC-2023668.